\documentstyle[12pt,aaspp4,epsf]{article}

\begin{document}

\def\intldate{\number\day\space\ifcase\month\or
January\or February\or March\or April\or May\or June\or
July\or August\or September\or October\or November\or December\fi
\space\number\year}

%
%

\def \event	{{EROS~2000-BLG-5}}
\def \shortname	{{EB2K005}}
\def \ncite  	#1{#1\ --{\it get ref}}
\def \deg    	{$^{\circ}$}
\def \etal   	{{et al.\thinspace}}
\def \eg     	{{e.g.,}}
\def \cf     	{{cf.}}
\def \ie     	{{i.e.,}}
\def \hub    	{$H_{\hbox{\rm 0}}$}
\def \hunits 	{km s$^{\hbox{\rm --1}}$ Mpc$^{\hbox{\rm --1}}$}
\def \kms    	{{\rm km~s$^{\hbox{\rm --1}}$}}
\def \sec    	{$^{s}$}
\def \arcsecpoint {{$^{''}\mkern-5mu.$}}
\def \dophot	{D{\sc o}PHOT}
\def \DOPHOT	{D{\sc o}PHOT}
\def \hi   	{\ion{H}{1}}
\def\sb		{{\rm mag~arcsec$^{-2}$}}
\def\area	{${\rm deg}^2$}
\def\kpc	{\hbox{\rm kpc }}
\def\pc		{\hbox{ pc }}
\def\yr		{ \, {\rm yr}}
\def\peryr	{ \, {\rm yr^{-1} }}
\def\vlos	{ v_{\rm los} }
\def\lsim	{ \rlap{\lower .5ex \hbox{$\sim$} }{\raise .4ex \hbox{$<$} } }
\def\gsim	{ \rlap{\lower .5ex \hbox{$\sim$} }{\raise .4ex \hbox{$>$} } }
\def\solar	{ {\odot} }
\def\lsolar	{ {\rm L_{\odot}} }
\def\msolar	{ \rm {M_{\odot}} }
\def\rsolar	{ \rm {R_{\odot}} }
\def\surfmunit  { \rm {\, \msolar \, pc^{-2}} }
\def\HI		{{H{\sc I}}}
\def\mags	{{ \, \rm mag }}   
\def\percubicpc	{ { \pc^{-3} } }
\def\abs	{ \hbox{ \vrule height .8em depth .4em width .6pt } \,} 
\def \tightenlines {\def\baselinestretch{1}\small}
\def\gtorder	{\mathrel{\raise.3ex\hbox{$>$}\mkern-14mu\lower0.6ex\hbox{$\sim$}}}
\def\ltorder	{\mathrel{\raise.3ex\hbox{$<$}\mkern-14mu\lower0.6ex\hbox{$\sim$}}}

%
%
%

\title{
H$\alpha$ Equivalent Width Variations across the 
Face of a\\ Microlensed K Giant in the Galactic Bulge\footnote{
Based on observations at the European Southern Observatory  
(Programs 265.C-5728 \& 265.C-5729)}}

\author{
M. Albrow\altaffilmark{2},
J. An\altaffilmark{3},
J.-P. Beaulieu\altaffilmark{4},
J. A. R. Caldwell\altaffilmark{5},
M. Dominik\altaffilmark{6}, \\
J. Greenhill\altaffilmark{7},
K. Hill\altaffilmark{7},
S. Kane\altaffilmark{7},
R. Martin\altaffilmark{8},
J. Menzies\altaffilmark{5},
K. Pollard\altaffilmark{9},\\
P. D. Sackett\altaffilmark{6},
K. C. Sahu\altaffilmark{2}, 
P. Vermaak\altaffilmark{4},  
R. Watson\altaffilmark{7},
A. Williams\altaffilmark{8} \\
(The PLANET Collaboration) \\
and 
P.H. Hauschildt\altaffilmark{10}}

\altaffiltext{2}{Space Telescope Science Institute, 3700 San Martin Drive, Baltimore, MD~21218, USA}
\altaffiltext{3}{Department of Astronomy, Ohio State University, 140 W.~18th Avenue, Columbus, OH~43210, USA}
\altaffiltext{4}{Institut d'Astrophysique de Paris, INSU CNRS, 98 bis Boulevard Arago, F-75014, Paris, France}
\altaffiltext{5}{South African Astronomical Observatory, P.O. Box 9, Observatory 7935, South Africa}
\altaffiltext{6}{Kapteyn Astronomical Institute, Postbus 800, 9700 AV Groningen, The Netherlands}
\altaffiltext{7}{Univ. of Tasmania, Physics Dept., G.P.O. 252C, Hobart, Tasmania~~7001, Australia}
\altaffiltext{8}{Perth Observatory, Walnut Road, Bickley, Perth~~6076, Australia}
\altaffiltext{9}{Physics Department, Gettysburg College, Gettysburg,
PA,~17325-1493, USA}
\altaffiltext{10}
{Department of Physics and Astronomy \& Center for Simulational Physics,
University of Georgia, Athens, \newline\hglue 1cm\ GA~30602-2451, USA}


\begin{abstract}
We present VLT FORS1 spectroscopy that temporally resolves the second
caustic crossing of the Bulge K giant source of
microlensing event \event, the first time this has been accomplished for
several phases of a caustic transit.
The $\sim$1 \AA\ H$\alpha$ equivalent width of the source star increases
slightly as the center of the star egresses the caustic and then
plummets by 30\% during the final limb crossing. These changes
are not seen in contemporaneous spectra of control stars in the FORS1 slit,
but are qualitatively consistent with expectations from stellar atmosphere
models as the caustic differentially magnifies different portions of
the stellar face of the target.  Observations such as these in a variety of 
stellar lines are equivalent to atmospheric tomography and are 
expected to provide a direct test of stellar models.
\end{abstract}

\vglue -0.1cm



\section{Introduction} \label{intro}

The steep magnification gradient near the
caustics\footnote{\footnotetext ~~A caustic is
a locus of positions in the source plane for
which the magnification is formally infinite.} 
generated by a binary microlens both magnifies and resolves
the background star as the sharply peaked lensing ``beam'' sweeps
across the source.  The finite size of the source and its wavelength-dependent
surface brightness profile leave their signatures
on the shape of the resulting microlensing light curve.  
The largest effect is the broadening and decrease in amplitude of the
caustic peaks in the light curve
which is related to the size of the source compared
to that of the Einstein ring and can be used to measure the relative proper
motion of the lens and source (Alcock \etal\ 1997, 2000;
Albrow \etal\ 1999a, 1999b, 2000, 2001a).  The next highest
order term is that due to broadband limb darkening: at most
wavelengths, the stellar disk is brighter at center than at limb,
which steepens the slope of the light curve when the caustic transits
the stellar limb.  These effects were predicted several years ago
(Schneider \& Wagoner 1987; Witt \& Mao 1994; Witt 1995;
Bogdanov \&  Cherepashchuk 1995; Gould \& Welch 1996)
and have now been used to measure wavelength dependent limb-darkening
coefficients for three Bulge giants (Albrow \etal\ 1999b, 2000, 2001a) and one
A dwarf in the Small Magellanic Cloud (Afonso \etal\ 2000).  Such measurements are
important to stellar atmosphere physics since very few techniques are 
available to extract reliable limb-darkening coefficients for typical
stars other than the Sun.

Time-resolved spectroscopic monitoring with large aperture telescopes
now offers the means to go one step further.  Measurements of the 
changes in the equivalent width and line shape of individual emission
and absorption
lines in the source star during a caustic crossing are now feasible
(Loeb \& Sasselov 1995; Valls-Gabaud 1998)
and can be expected to yield detailed information about
the chemothermal structure of the atmosphere as a function of
depth (Heyrovsk\'y, Sasselov \& Loeb 2000).
Starspots and the polarization and kinetic structure of the stellar
envelopes may also be detectable with the aid of microlensing caustics
(Simmons, Newson \& Willis 1995; Simmons, Willis \& Newsam 1995;
Igance \& Hendry 1999, Bryce \& Hendry 2000, Heyrovsk\'y \& Sasselov 2000).

Fold (line) caustics, such as the one that transited the source star of \event,
would be expected to produce somewhat smaller differential signals
($\sim$1\%) than direct
transits by (single-lens) point caustics (Gaudi \& Gould 1999)
because they have a broader ``beam pattern''.
On the other hand, fold caustic transits are observed much more frequently
than are direct point transits.  Detection of the
differential signal in individual spectral lines thus requires not
only large apertures and efficient spectrographs to achieve the high S/N, but also
control or understanding of systematics at the $\sim$1\% level.

Following a real-time electronic microlensing alert issued by the EROS group, 
the PLANET collaboration (Albrow \etal\ 1998)
commenced intense photometric monitoring
of the event \event.  A secondary alert issued by the MPS collaboration
announced an unexpected, sudden, and significant brightening of the object;
PLANET increased its sampling rate immediately.  The resulting dense
coverage allowed characterization of the first caustic crossing
of this binary event by a technique advocated by Albrow \etal\ (1999c).
Together with subsequent high precision photometry in the caustic trough,
a prediction for the timing of the second caustic crossing could then be made,
which was announced electronically by
PLANET\footnote{\footnotetext ~~See http://www.astro.rug.nl/$\sim$planet/EB2K005.html}
a few days in advance, facilitating the VLT observations reported here.
The exceptionally long duration of this crossing (due to the glancing angle at
which the source left the caustic region) allowed
PLANET spectroscopic monitoring from Paranal
to be spread over several nights, during which time the source was
differentially resolved as the
caustic passed over the center and then the limb of the exiting source.

In this Letter, we describe the changes in the equivalent width of the H$\alpha$
line observed during our VLT FORS1 spectroscopic monitoring of the K-giant 
source star of \event, 
the first time that full temporal coverage has been obtained of a
caustic crossing at such high spectral resolution.  We show that these
changes in H$\alpha$ are qualitatively consistent with
those expected from stellar atmosphere models for a star of this type;
more detailed comparisons with theory over the complete spectrum
will be presented elsewhere (Albrow \etal\ 2001b).

\section{FORS1 Observations of \event}\label{data}

The $\sim$four-day duration of the second caustic crossing of \event\ allowed PLANET to
spread its allocation of 9.5 hours of Director's discretionary time on the VLT 
over several nights; one pre-caustic night when the source star was magnified, but unresolved,
and about two hours observation on each of four nights during
the caustic crossing.  Figure~1 shows the times at which the spectra
were obtained relative to model light curves; other
details are given in Table~1. The FORS1 spectrograph on UT1 (Antu)
was used with the atmospheric dispersion compensator (ADC) and a long slit of
projected width 1\arcsec\ ($\sim$5~pixels).
The 600R ($5380 - 7530$~\AA) and 600I ($7050 - 9180$~\AA) gratings
provided a dispersion of $\sim$1 \AA\  per pixel,
so that in periods of 0.6\arcsec\ seeing the spectral resolution was $\sim$3\AA.
On the first night, spectra were also taken with the 600B ($3900 - 5800$~\AA) grism.
At peak, the source reached $I_C \sim 12.6$~mag.
As a means of studying systematic effects, 
the slit was oriented to include other stars in the field with
brightness and spectral type comparable to those of the target.
By design, most spectra were taken at similar airmasses ($\lsim 1.2$).

We used four-amplifier readout of the CCD, for which the ESO pipeline is not yet 
available.  The raw images were flatfielded, and 
spectra extracted and wavelength calibrated in the usual way within the
IRAF\footnote{~~IRAF is distributed by the National Optical Astronomy
Observatories, which are operated by the Association of Universities
for Research in Astronomy, Inc., under cooperative agreement with the
National Science Foundation.} environment.
The rms error of the wavelength fits was typically 0.08 -- 0.10 \AA.  
Attempts to improve this using night sky emission lines did not result in
significantly better accuracy.  All nights appeared to be photometric.
Flux calibration was effected with reference to spectra of LTT377 and LTT3218
that were obtained on the first night.
A representative spectrum, extended blueward with observations
from the first night, is shown in Figure~2.
Modeling of the continuum and relative
line strengths with a grid of Kurucz (1992) model fluxes
indicates that the source star is probably a heavily reddened K3 giant with
$T_{\rm eff} = 4500 \pm 250$K, log~$g = 3.5 \pm 0.5$, $E(B-V) = 1.30 \pm 0.05$,
and [Fe/H] $= -0.3 \pm 0.1$, where the uncertainties reflect
the step size in the fitting grid.   A color-magnitude diagram of the
field indicates that the source color and brightness are consistent
with a K giant in the Bulge.

Spectra extracted for seven stars other than the target
provide a useful check on systematic effects from night to night.
In particular, three stars that we designate
S3, S4, and S6 are fairly well centered on the slit for
all spectra and are used here as comparison stars.
Star S6 is a bright M giant; S3 and S4 are K giants of brightness
similar to that of the unamplified target.
Although they show no large differences with respect to other frames,
we exclude from our final analysis two spectra taken on the first night in airmasses
corresponding to zenith angles larger than the recommended limit (50\deg ) of the ADC.

\section{Measuring Temporal Changes in the H$\alpha$ Equivalent Width}

Two independent methods were employed to search for significant
changes in line strengths during the caustic crossing of \event.
The first method used cross correlation techniques to adjust (at subpixel levels)
the wavelength calibration of the spectra.  The spectra from a
given night were then coadded and used to form spectral ratios
of one night over another.   In the second approach, the equivalent
widths of lines were measured by computing the area under a line connecting
two points that lay close to the continuum flux of the star on either side
of the spectral feature.  This method is independent of slight changes in
the wavelength calibration from night to night, but may depend on the choice of
the continuum points. Both approaches indicate a significant temporal signal of
comparable size in the equivalent width of
the H$\alpha$ line of the target.
The control stars showed no or very
much smaller marginal signatures ($\lsim \, 2\sigma$) when analyzed in the same way.

The nightly ratios resulting from the spectral ratio analysis
are shown in Figure~3 for the source star and for comparison stars S3, S4 and S6.
The target shows a clear nightly change in equivalent width.
More flux is present at H$\alpha$ (i.e., the absorption line is
less deep) on Night 5 than on any another night; the difference between
Nights 3 and 5 is the greatest.  Ratios formed from Night 4
rather than Night 5 show the same trend, confirming that the strong temporal signal seen
in the target is not an artifact in the data of the last night of observation.
Differences in equivalent width relative
to Night 5 are given in Table~1.

In Figure~4, the directly-determined H$\alpha$ equivalent widths of the
target are plotted as a function of the Julian date at mid-exposure.
Nightly averages are shown with error bars reflecting the standard error
in the mean of the measured values.
The equivalent width of the target increases 
slightly as the (hotter) center of the star passes over the caustic and
clearly decreases as the trailing limb exits, a difference of
30~$\pm 3$\% is measured between these two nights.
These trends are seen in individual spectra as well as in the nightly
averages; no single spectrum or subgroup of spectra dominates the signal.
Within the uncertainties, the decrease in H$\alpha$ equivalent width between
Nights 3 and 5 found in the spectral ratio analysis is the same
as that found by direct computation of the equivalent widths (Table~1).

\section{Comparison to a K Giant Atmosphere Model}

We conclude that the changes in H$\alpha$ equivalent width observed by FORS1
in the K giant source of \event\ are due to spatial resolution of
the star's atmosphere during the caustic crossing, and proceed to compare them with 
expectations from atmospheric models.
For the demonstration purposes in this Letter, we choose one example
roughly matching the spectral characterization of our target
(solar metallicity, $M = 1 M_\odot$, ${\rm log~} g = 3.5$ and
$T_{\rm eff} = 4400$K) from a set of high-resolution, spherically-symmetric
model atmospheres for giants
(Hauschildt \etal\ 1999; Orosz \& Hauschildt 2000).
Spectra derived from differing annuli across
the face of such a star vary rather dramatically: those near the limb are
generally redder and exhibit emission line features that are in absorption
near the center of the star.  The caustic samples these differences
as it sweeps across the source. 

We model the magnification pattern across the face of the microlensing source
as that given by rectilinear motion across a simple fold caustic (with
constant background magnification).
Near such a fold and inside the caustic structure
itself, the magnification grows as $x^{-1/2}$, where $x$ is the perpendicular 
distance from the source element to the fold caustic.
The height and width of the resulting light curve
are adjusted to match, as closely as possible within these models,
our independent photometric data over the second crossing of this event.
This determines the mapping from $x$ to time, so that the timing of
our FORS1 spectra can be related to a specific beam magnification pattern
over the source.
Integrating these changing magnification patterns over the
model atmosphere for all wavelengths near H$\alpha$ allows us to compute
the evolution of the H$\alpha$ line expected during this
caustic crossing.

In particular, we have calculated the temporal change in the
equivalent width for two simple fold models with differing caustic strength
and background magnification (Fig.~1) that mimic the general features
of the observed light curve on Nights 2 though 5.
The results are plotted over our observations in Figure~4.
The two dips in the models mark the times
at which the leading and trailing limbs exit the caustic.
The only free parameter is an overall scale factor of 0.81 applied to
match the equivalent width of the fiducial model to that of the unresolved source.
These simple models agree qualitatively with the spectral data,
producing quite similar equivalent widths for all nights but the last.
The model with the higher background magnification --
more similar in this respect to the observed light curve --
produces the larger equivalent width on Night~5.
However, the observed light curve also exhibits a feature
not modeled by any fold caustic: a fast rise
to large magnification directly after Night~5 that is characteristic
of the source trajectory passing near a caustic cusp.
More physically-realistic models are thus needed
before a more quantitative comparison with stellar models can be made.

The CaII triplet would also be expected to exhibit changes in equivalent
width during the caustic transit.
We see indications of such changes in our FORS1 spectra,
but narrowness of the lines and contamination from
sky emission complicates the analysis of the Ca~II equivalent
widths; these will be presented elsewhere (Albrow \etal\ 2001b).

\section{Discussion}

Center-to-limb variations are expected in many strong absorption lines of cool giants,
including H$\alpha$.  Such lines are expected
to be deeper just before and during the egress of the center of a giant
microlensed star from a fold caustic.  Since the line is in emission
at the coolest portions of the outer atmosphere, 
the observed line should become shallower (smaller equivalent width)
as the limb egresses and becomes magnified differentially with respect to the rest
of the star.  This behavior is confirmed by our observations
of \event\ throughout the final transit of the K giant source
star by the caustic.

Previous attempts to measure spectral line changes during
caustic crossings have been
encouraging, but hampered by the lack of temporal coverage during those
phases of the caustic transit that most significantly resolve source structure.
The real-time alert provided by the MACHO team
(Alcock \etal\ 2000) allowed Lennon \etal\ (1996) to take NTT spectra
over the peak of the fold crossing in MACHO~96-BLG-3,
but these did not extend far enough temporally to detect
spectral differences due to source resolution.
A difference in H$\alpha$ equivalent width during and after the
grazing point-caustic transit of the M giant source of
event MACHO~95-BLG-30 was reported by Alcock et al.\ (1997),
and appears to be qualitatively compatible with 1-D stellar models
(Sasselov 1998).  The reported changes in TiO equivalent width
for this event (Alcock \etal\ 1997) are difficult to interpret as they are
as large in the several days after the crossing as
during and just after the transit.

Our spectroscopic monitoring of \event\ with FORS1 on the VLT
is the first to be carried
out through all phases of a microlensing caustic crossing.
Significant changes in the equivalent width of H$\alpha$ were detected
throughout the transit.  A fiducial K giant model atmosphere
convolved with simple fold caustic magnification models
exhibits changes in equivalent width that are in qualitative agreement
with the observed changes.
At the time of this writing, PLANET continues
photometric monitoring of \event\ which, although far past the caustic structure,
is still declining in brightness.  Modeling indicates that
the source star passed quite close to a caustic cusp during the second
crossing, altering the temporal and spatial
behavior of the microlensing magnification ``beam''.
A quantitative confrontation with atmospheric models requires proper
treatment of the effect of the cusp over the whole stellar disk, and is
now in progress (Albrow \etal\ 2001b).

Stellar tomography using microlensing caustics is now feasible; adequate
alert mechanisms combined with modeling and dense, precise
follow-up photometry enable prediction of caustic transits days in advance.
Spectra of the source and field control stars of \event\
demonstrate that equivalent width changes as small as 0.1\AA\
and relative spectral differences of $\sim$1\% per
resolution element can be observed
in two hours (per temporal sampling element)
with FORS on the VLT for microlensed Bulge sources.
These characteristics are well-matched to monitoring the
equivalent width of many lines during a (typical) 8-24 hour
caustic transit.  The evolution of spectral line shape, an
even more sensitive indicator of stellar atmosphere conditions, will require the
higher resolution provided by Echelle instruments.
If large aperture telescopes with such spectroscopic ability, especially
those in the southern hemisphere, can be flexibly scheduled in real-time,
microlensing can meaningfully test theories of
stellar atmospheres in the next few years.


\acknowledgments
PLANET thanks EROS for the public real-time alert of \event\
that allowed us to begin photometric monitoring
and MPS for the anomaly alert that resulted in our increased
sampling over the first caustic.
We also thank Malcolm Hartley, Quentin Parker,
and Tom Lloyd Evans, for taking spectra at MSSSO and SAAO that are
not included in this analysis.  PLANET is especially grateful to the
ESO directorship for awarding discretionary time and to
the ESO scientific and technical staff
for their skilled, unflagging and cheerful help during these
necessarily hurried service observations.
This work was supported by award GBE 614-21-009
from the Dutch Organization for Scientific Research (NWO) and by a
donation to the University of Tasmania by Mr.~David Warren.


\noindent{\bf Note Added in Proof:~}
After preprint circulation of this work
(astro-ph/0011380) and receipt of the referee's report of our submitted
manuscript, we became aware of the work of Castro et~al.
(astro-ph/0101025) from Nights 3 and 4, which confirms our results.





\begin{table}
\caption{Log of R-band Spectroscopic Observations of \event\
\label{logtab}}
\begin{tabular}{cccccc}
 & & & & & \\
~~Date &  No.~of  & Typical & Seeing   & H$\alpha$ Equiv.~Width (in $\AA$)\\
     &       Spectra & Exp.~time &  FWHM   & Method 1~~~~~Method 2  \\
     &               &   (sec) & (arcsec) & (Relative)~~~~~(Absolute)\\
     \\
25-Jun-00 & 5  & 300 & $0.8 - 1.5$   & 0.27 $\pm$ 0.04~~~1.03 $\pm$ 0.04 \\
~4-Jul-00 & 10 & 360 & $0.6 - 0.75$ & 0.25 $\pm$ 0.03~~~1.06 $\pm$ 0.01 \\
~5-Jul-00 & 10 & 360 & $0.5 - 0.75$ & 0.28 $\pm$ 0.03~~~1.09 $\pm$ 0.02 \\
~6-Jul-00 & 12 & 540 & $0.6$        & 0.17 $\pm$ 0.02~~~0.98 $\pm$ 0.01 \\
~7-Jul-00 & 4 & 540 & $0.6$         & ~~~~~~~---~~~~~~~~0.79 $\pm$ 0.03 \\
\end{tabular}
\end{table}
\clearpage

\section*{Figure Captions}


\figcaption{
Simple fold caustic model light curves
of \event\ showing the phase intervals, marked by pairs of arrows,
during which the caustic-crossing spectra discussed here were obtained.
The peak of the light curve occurs before the time at which
the center of the star exits the caustic, which in these models
is indicated by the dashed line.
The trailing limb exits near JD = 2451733.7.
}
\figcaption{
A representative spectrum of \event\ taken with FORS1 on the VLT.
The insert shows a 100 \AA\ region around the H$\alpha$ line using the same
vertical scale.  Dereddening has not been applied.
}

\figcaption{
Ratios of the average flux measured on two different nights
shown as a function of wavelength in the region around H$\alpha$.
Top panel is for the target, lower panels for comparison stars
S3, S4, and S6.  The ratios are (from top to bottom):
Night~5/Night~1; Night~5/Night~2; Night~5/Night~3; Night~5/Night~4;
and Night~4/Night~3.
Gaussian smoothing of 3 pixels (FWHM) was performed before the ratios were formed,
which were then divided by their mean (indicated by the dashed line) before display.
Results are offset from one another for clarity.
The first ratio is the noisiest primarily due to poor seeing on the first night.
}

\figcaption{
The equivalent width of H$\alpha$ for the source star of 
\event\ as a function of
Julian date at mid-exposure. Nightly means are shown as the larger filled dots. 
The broken line passes through the mean value of the first night.
Also shown (solid lines) are the
predicted relative changes in equivalent width for a fiducial model
atmosphere transited by two different simple fold caustic
models (see Fig.~1 and \S4).
}

\newpage

\vglue 1cm
\hglue -1cm
\epsfxsize=15cm\epsffile{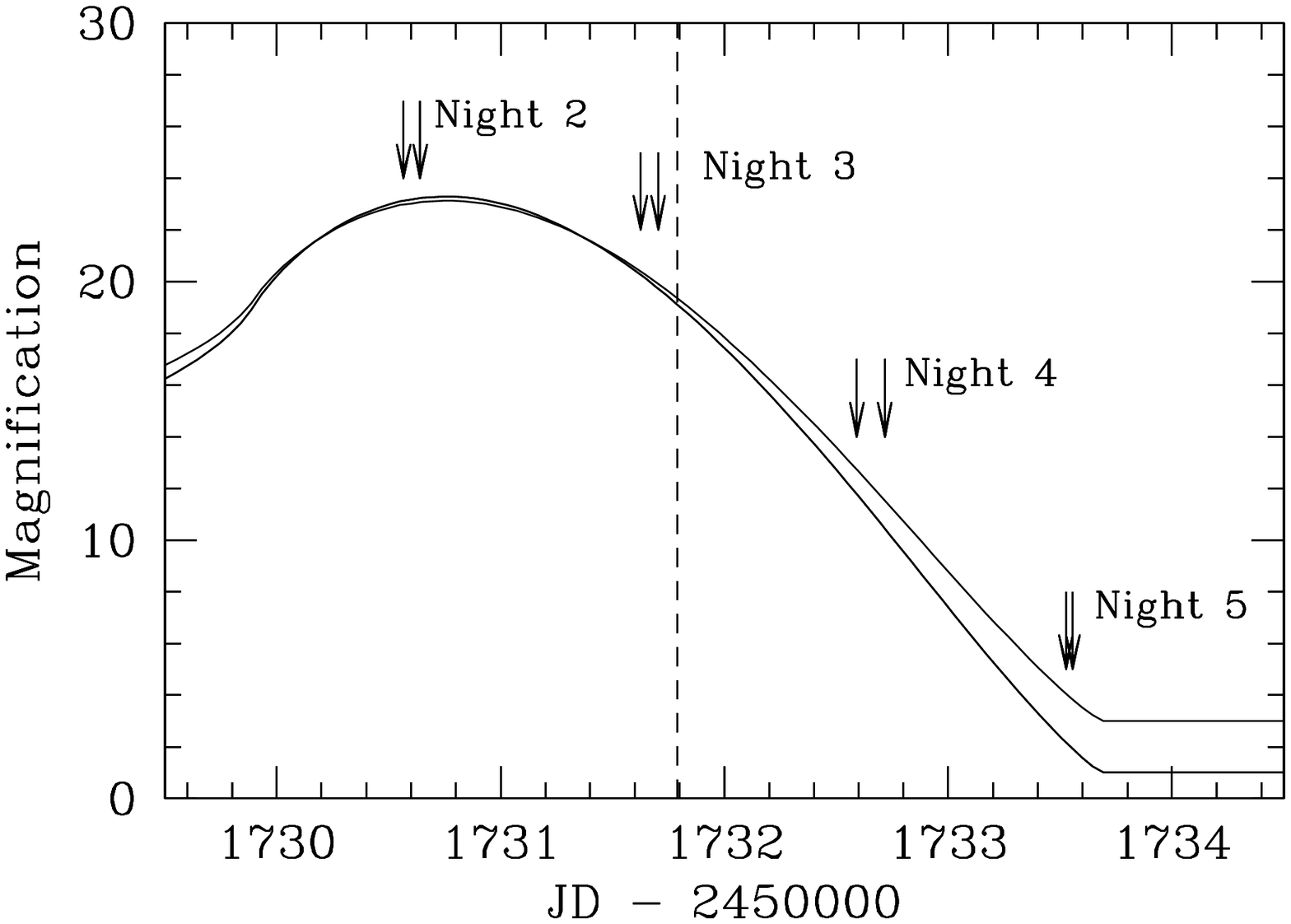}

\newpage

\vglue -5cm
\hglue -2cm
\epsfxsize=18cm\epsffile{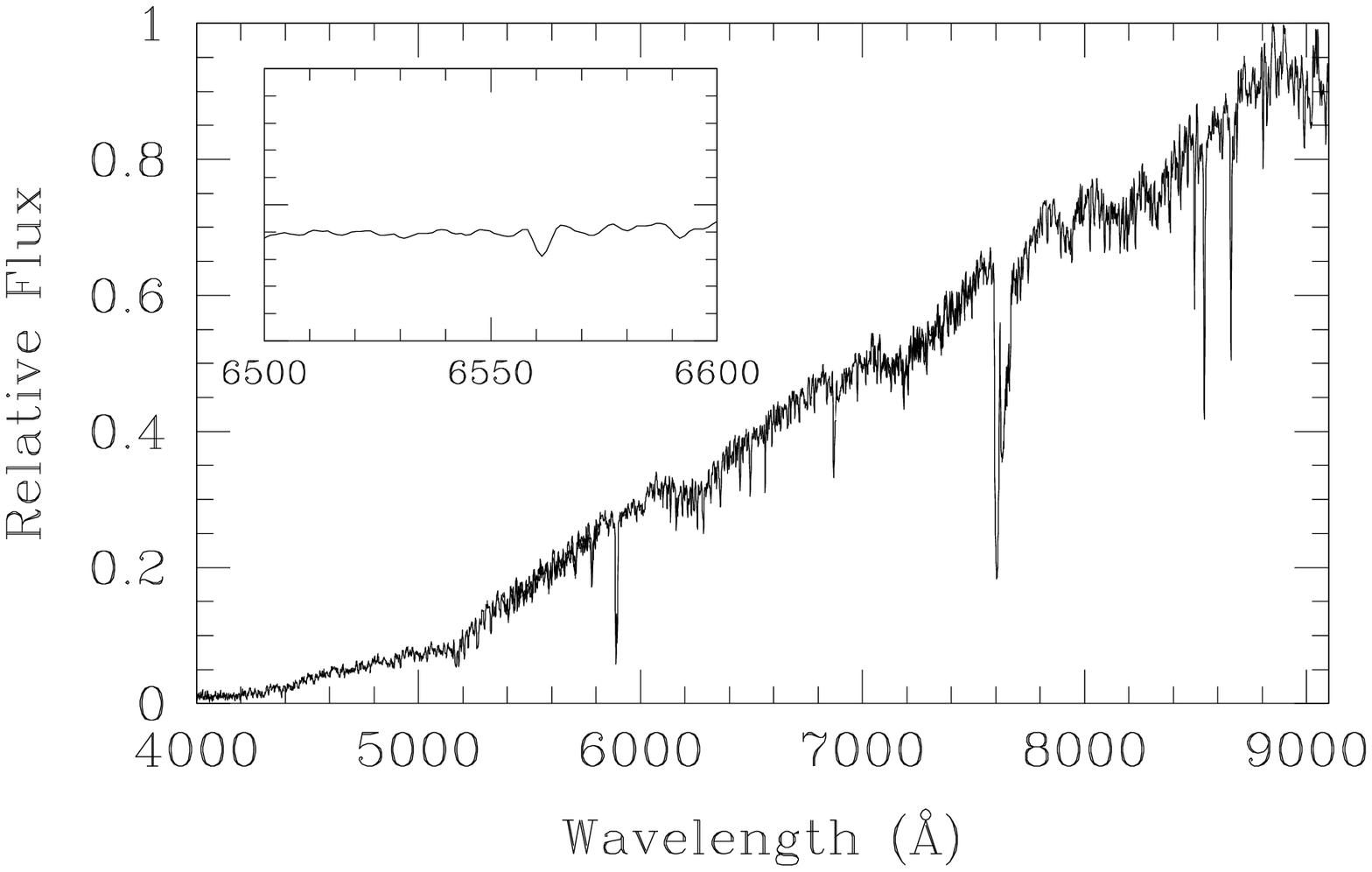}

\newpage

\hglue -5cm
\epsfxsize=24cm\epsffile{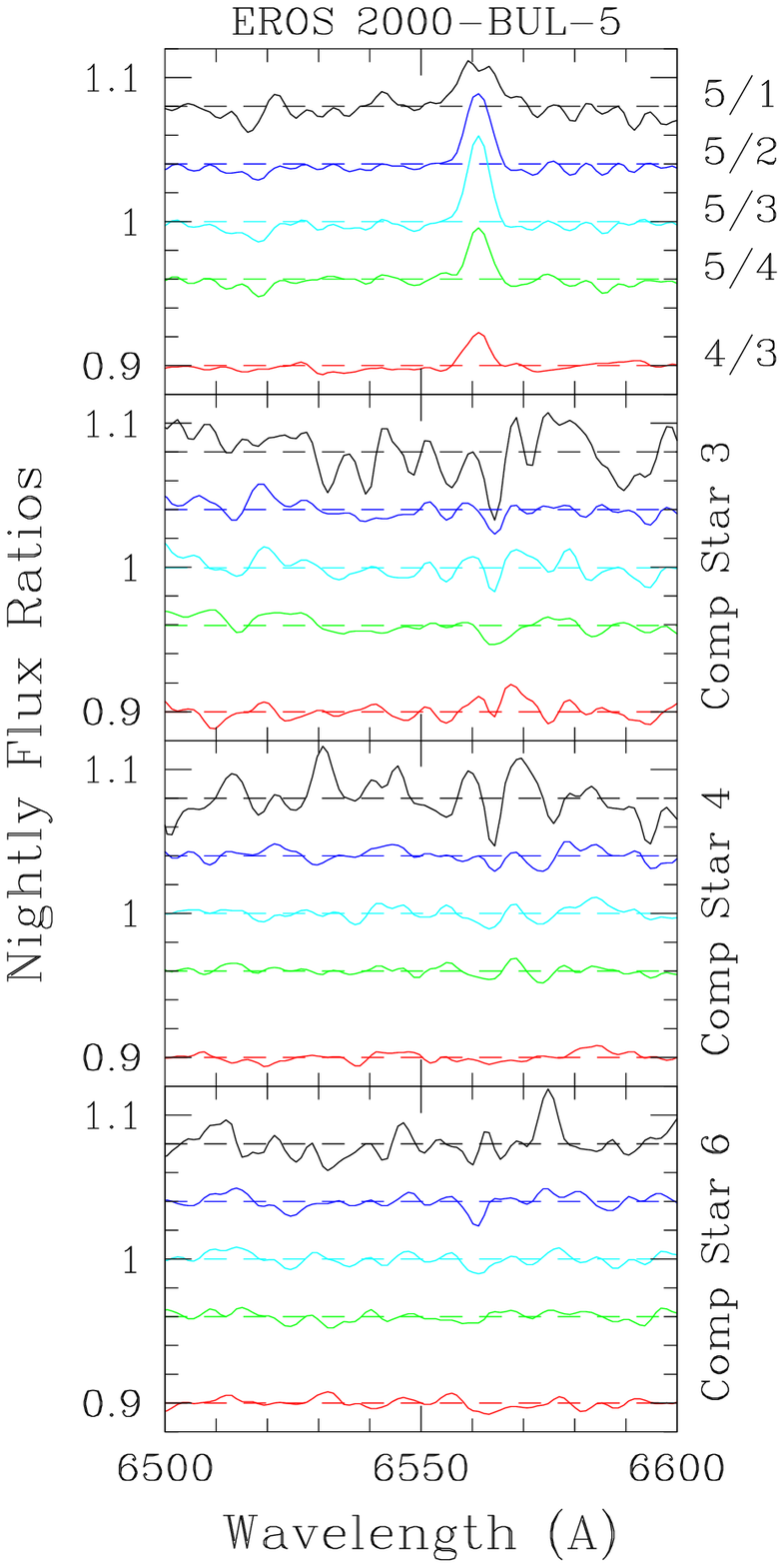}

\newpage

\vglue -2cm
\hglue -2cm
\epsfxsize=18cm\epsffile{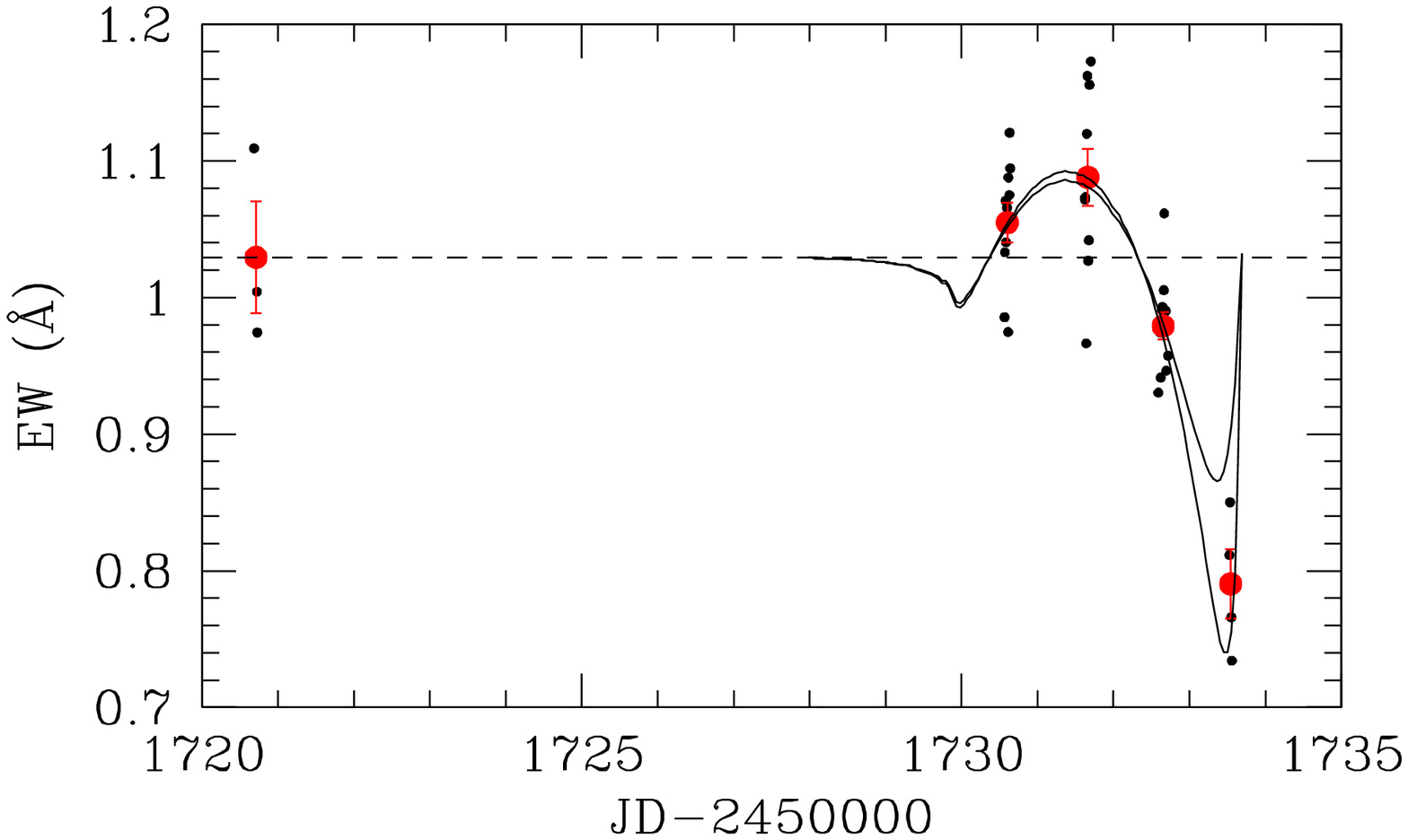}

\end{document}